\documentstyle[aps]{revtex}

\begin{document}
\author{F. T. Vasko}
\address{Institute of Semiconductor Physics, National Academy of Sciences, 45 Pr.\\
Nauky, Kiev, 252650, Ukraine}
\author{O. G. Balev$^{a}$ and Nelson Studart}
\address{Departmento de F\'{i}sica, Universidade Federal de S\~{a}o Carlos,\\
13565-905, S\~{a}o Carlos, S\~{a}o Paulo, Brazil}
\date{June 16, 2000}
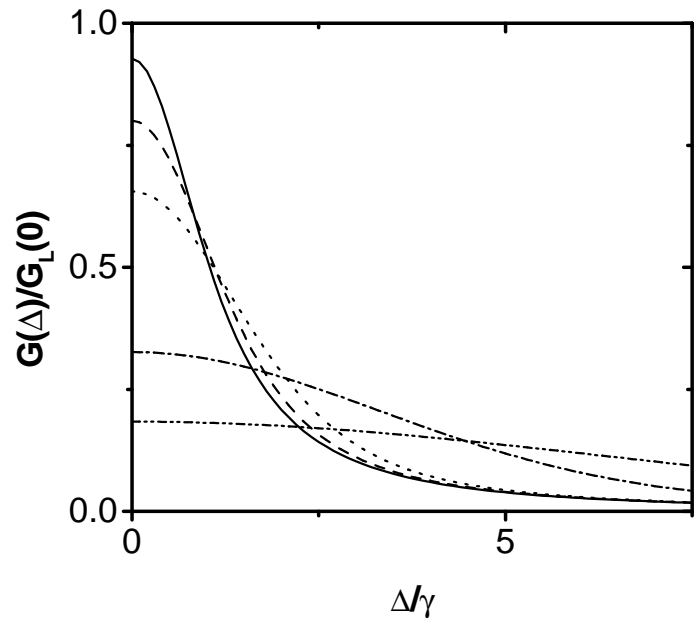
\title{Inhomogeneous broadening of tunneling conductance in double quantum wells }
\maketitle

\begin{abstract}
The lineshape of the tunneling conductance in double quantum wells with a
large-scale roughness of heterointerfaces is investigated. Large-scale
variations of coupled energy levels and scattering due to the short-range
potential are taken into account. The interplay between the inhomogeneous
broadening, induced by the non-screened part of large-scale potential, and
the homogeneous broadening due to the scattering by short-range potentials
is considered. It is shown that the large inhomogeneous broadening can be
strongly modified by nonlocal effects involved in the proposed mechanism of
inhomogeneity. Related change of lineshape of the resonant tunneling
conductance between Gaussian and Lorentzian peaks is described. The
theoretical results agree quite well with experimental data.
\end{abstract}


\section{Introduction}

Resonant tunneling in semiconductor heterostructures has been widely
investigated ever since Tsu and Esaki proposed the double-barrier
resonant-tunneling diode{\cite{1}} (see Ref. \cite{2} for a recent review).
New developments came through from studies of interlayer tunneling
spectroscopy between parallel two-dimensional electron systems (2DES) using
the technique of independent contacts to closely located 2DES. \cite{3,4}.
The 2DES are formed in two GaAs quantum wells (QW) separated by a Al$_{x}$Ga$%
_{1-x}$As barrier. Because the in-plane momentum and the energy are
conserved, the 2D-2D tunneling current exhibits sharp resonance peak whose
broadening is determined by different collision processes in the nonideal
double quantum well (DQW) structure. This property allows to study
scattering mechanisms through tunneling spectroscopy method.\cite{5}
Furthermore, broadening effects may be important in a novel quantum
transistor based on 2D-2D tunneling in independently contacted DQWs.\cite{6}

The aim of this paper is to describe the lineshape of the resonant tunneling
current in nonideal DQWs with independent contacts to each QW, when, in
addition to usual homogeneous broadening induced by short-range scattering,
the inhomogeneous broadening due to large-scale variations of
heterointerfaces is taken into account. The latter scattering mechanism has
an essential effect on the form of the peak, because smooth variations of
the DQW energy levels due to large-scale random variations of the widths of
right ($r$-) and left ($l$-) QWs can not be screened, even though the
screening potential involves all possible redistributions of electrons
within the DQW structure. Even though the averaged large-scale potential is
screened in heavily doped structures, the intersubband energy is still
nonuniform over the plane of the quantum well. In Fig. 1, a schematic view
of the band diagram of DQWs and spatial variations of the energy levels are
depicted for illustration. Our theory is valid when the DQW width is smaller
than the correlation length $\ell _{c}$ for nonuniformities of the
heterointerfaces in the DQW. A very similar mechanism was recently proposed
in a single quantum well for describing the inhomogeneous broadening of
intersubband transitions, with one subband occupancy, \cite{7} and for new
effects in classical magnetotransport in the case of double subband
occupancy.\ref{8}

We show that the Lorentzian lineshape for the tunneling current peak, in the
case of short-range collision-induced broadening, assumes a Gaussian shape
due to the inhomogeneous broadening, if nonlocal effects are discarded due
to sufficiently large $\ell _{c}$. However, for not too large $\ell _{c},$
we obtain the transformation from a Gaussian to Lorentzian lineshape due to
nonlocal effects on the inhomogeneous broadening. Moreover, inhomogeneous
and nonlocal effects essentially modify the half width at half maximum
(HWHM) of the peak. As it is shown below, our theoretical results are in
quite reasonable agreement with the experimental ones of Ref. \cite{5}.

The paper is organized in a following way. In Sec. II we evaluate the
expression for the tunneling current up to second order in the weak
interwell tunneling coupling and use the path-integral representation to
calculate the tunneling conductance in terms of the averaged product of
Green's functions for electron in left ($l$-) and right ($r$-) QWs. The
lineshape of the resonant tunneling conductance is analyzed in Sec. III in a
quasiclassical approximation. The list of assumptions and concluding remarks
are given in the Sec. IV. The Appendix A contains estimates of the
parameters used in the nonscreened potential, due to large scale
nonuniformities of the heterointerfaces, and in Appendix B we briefly
discuss the optimal fluctuation method and the straightforward trajectory
approximation used in Sec. III.

\section{Tunneling current}

Electron states in $l$- and $r$-QWs are described by the Hamiltonians 
\begin{eqnarray}
\hat{H}_{l} &=&\Delta +\frac{\hat{p}^{2}}{2m}+\overline{U}_{l{\bf x}}+%
\widetilde{U}_{l{\bf x}}+V_{{\bf x}},  \nonumber \\
\hat{H}_{r} &=&\frac{\hat{p}^{2}}{2m}+\overline{U}_{r{\bf x}}+\widetilde{U}%
_{r{\bf x}}+V_{{\bf x}},  \label{1}
\end{eqnarray}
where $\Delta $ is the interlevel splitting without tunneling and $m$ is the
effective mass. The effect of fluctuations of heterointerfaces and
scattering processes are described by large-scale and short-range potentials 
$\overline{U}_{l,r{\bf x}}$ and $\widetilde{U}_{l,r{\bf x}}$ in $l$, $r$%
-QWs. The screening potential, $V_{{\bf x}}$, included in $\hat{H}_{l,r}$,
is determined from the Poisson equation (see Appendix A) and only the
averaged large-scale potential is screened as

\begin{equation}
\frac{\overline{U}_{l{\bf x}}+\overline{U}_{r{\bf x}}}{2}+V_{{\bf x}}=0.
\label{2}
\end{equation}
Taking into account the interwell tunneling coupling, we use a $2\times 2$
one-electron Hamiltonian matrix as

\begin{eqnarray}
&&\left| 
\begin{array}{ll}
\hat{h}_{l} & T \\ 
T & \hat{h}_{r}
\end{array}
\right| ,  \nonumber  \label{3} \\
&&
\end{eqnarray}
where the diagonal terms are given by 
\begin{eqnarray}
\hat{h}_{l} &=&\Delta +\frac{\hat{p}^{2}}{2m}+\widetilde{U}_{l{\bf x}}+\frac{%
\delta U_{{\bf x}}}{2},  \nonumber \\
\hat{h}_{r} &=&\frac{\hat{p}^{2}}{2m}+\widetilde{U}_{r{\bf x}}-\frac{\delta
U_{{\bf x}}}{2},  \nonumber
\end{eqnarray}
the non-screening part of the large-scale potential is $\delta U_{{\bf x}}=%
\overline{U}_{l{\bf x}}-\overline{U}_{r{\bf x}}$ and the nondiagonal terms
are given by the tunneling matrix element $T$ (the coupling energy). In the
following, we assume that the random potentials introduced above are
statistically independent and described by Gaussian correlation functions 
\begin{eqnarray}
\langle \widetilde{U}_{j{\bf x}}\widetilde{U}_{j^{\prime }{\bf x}}\rangle
&=&\delta _{{jj}}{^{\prime }}\widetilde{W}_{j}(|{\bf x}-{\bf x^{\prime }}%
|),\qquad \langle \overline{U}_{j{\bf x}}\overline{U}_{j^{\prime }{\bf x}%
}\rangle =\delta _{{jj}}{^{\prime }}\overline{W}_{j}(|{\bf x}-{\bf x^{\prime
}}|),  \nonumber \\
&&\langle \delta U_{{\bf x}}\delta U_{{\bf x^{\prime }}}\rangle =\overline{W}%
_{l}(|{\bf x}-{\bf x^{\prime }}|)+\overline{W}_{r}(|{\bf x}-{\bf x^{\prime }}%
|)\equiv w(|{\bf x}-{\bf x^{\prime }}|),  \label{4}
\end{eqnarray}
where the functions $\overline{W}_{l,r}({\bf x})$ and $w({\bf x})$ are
discussed in Appendix A. We also neglect here the in-plane variations of the
matrix element $T$ (see discussion in \ Ref. \cite{9}).

The interwell tunneling current is expressed in terms of the density matrix $%
\hat{\rho}_{t}$ according to \cite{9,10} 
\begin{equation}
J_{\bot }=\frac{|e|T}{\hbar }\frac{2}{L^{2}}{\rm tr(}\hat{\sigma}_{y}\hat{%
\rho}_{t}),~\text{where}~\hat{\rho}_{t}=\left| 
\begin{array}{ll}
\hat{\rho}_{lt} & \tilde{\rho}_{t} \\ 
\tilde{\rho}_{t}^{+} & \hat{\rho}_{rt}
\end{array}
\right| ,  \label{5}
\end{equation}
with $\hat{\sigma}_{y}$ being the $y$ component of the Pauli matrix and the
trace includes both the average over large-scale and short-range random
potentials and the summation over electron states. Non-diagonal and diagonal
components of the density matrix in Eq. (\ref{5}) are connected by the
relation ($\delta \rightarrow +0$) 
\begin{equation}
\tilde{\rho}_{t}=\frac{iT}{\hbar }\int_{-\infty }^{t}dt^{\prime }e^{\delta
t^{\prime }}e^{-i\hat{h}_{l}(t-t^{\prime })/\hbar }(\hat{\rho}_{lt^{\prime
}}-\hat{\rho}_{rt^{\prime }})e^{i\hat{h}_{r}(t-t^{\prime })/\hbar }.
\label{6}
\end{equation}
Using a set of wave functions $\left\langle {\bf x}|j\lambda \right\rangle
\equiv \psi _{j{\bf x}}^{\lambda }$ which are determined by the eigenvalue
problems in the $j$th QW $\hat{h}_{j}\psi _{j{\bf x}}^{\lambda }=\varepsilon
_{\lambda }\psi _{j{\bf x}}^{\lambda }$ we rewrite the tunneling current (%
\ref{5}) as 
\begin{equation}
J_{\bot }=i\frac{|e|T}{\hbar }\frac{2}{L^{2}}\left\langle \left\langle
\sum_{\lambda }[(r\lambda |\tilde{\rho}_{t}|r\lambda )-(l\lambda |\tilde{\rho%
}_{t}^{+}|l\lambda )]\right\rangle \right\rangle .  \label{7}
\end{equation}
Here $\langle \langle \ldots \rangle \rangle $ means the average over
short-range and large-scale potentials. After substitution of Eq. (\ref{6})
in Eq. (\ref{7}), we obtain 
\begin{eqnarray}
J_{\bot } &=&\frac{2\pi |e|T^{2}}{\hbar }\frac{2}{L^{2}}\left\langle
\left\langle \sum_{\lambda \lambda ^{\prime }}|(r\lambda |l\lambda ^{\prime
})|^{2}\delta (\varepsilon _{r\lambda }-\varepsilon _{l\lambda ^{\prime
}})(f_{r\lambda }-f_{l\lambda ^{\prime }})\right\rangle \right\rangle 
\nonumber \\
&=&\frac{2\pi |e|T^{2}}{\hbar }\frac{2}{L^{2}}\int_{\varepsilon _{{%
\scriptscriptstyle F}l}}^{\varepsilon _{{\scriptscriptstyle F}%
r}}d\varepsilon \left\langle \left\langle \sum_{\lambda \lambda ^{\prime
}}|(r\lambda |l\lambda ^{\prime })|^{2}\delta (\varepsilon _{r\lambda
}-\varepsilon )\delta (\varepsilon -\varepsilon _{l\lambda ^{\prime
}})\right\rangle \right\rangle \;,  \label{8}
\end{eqnarray}
where the above second equation is written for the zero-temperature case and 
$\varepsilon _{{\scriptscriptstyle F}j}$ is the quasi-Fermi level in the $j$%
th QW.

In order to calculate $J_{\bot }$, it is convenient to use the retarded ($R$%
) Green's functions for the electron in $l$- and $r$-QWs, which are defined
as 
\begin{equation}
{\cal G}_{j\varepsilon }^{\scriptscriptstyle R}({\bf x},{\bf x^{\prime }}%
)=\sum_{\lambda }\frac{\psi _{j{\bf x^{\prime }}}^{\lambda \ast }\psi _{j%
{\bf x}}^{\lambda }}{(\varepsilon _{j\lambda }-\varepsilon -i\delta )},
\label{9}
\end{equation}
and the advanced ($A$) Green's functions given by ${\cal G}_{j\varepsilon }^{%
\scriptscriptstyle A}({\bf x},{\bf x^{\prime }})={\cal G}_{j\varepsilon }^{%
\scriptscriptstyle R}({\bf x^{\prime }},{\bf x})^{\ast }$. The tunneling
current assumes the form 
\begin{equation}
J_{\bot }=\frac{|e|T^{2}}{2\pi \hbar }\frac{2}{L^{2}}\int_{\varepsilon _{{%
\scriptscriptstyle F}l}}^{\varepsilon _{{\scriptscriptstyle F}%
r}}d\varepsilon \int d{\bf x}\int d{\bf x^{\prime }}\sum_{ab={%
\scriptscriptstyle RA}}(-1)^{k}\left\langle \left\langle {\cal G}%
_{l\varepsilon }^{a}({\bf x},{\bf x^{\prime }}){\cal G}_{r\varepsilon }^{b}(%
{\bf x^{\prime }},{\bf x})\right\rangle \right\rangle ,  \label{10}
\end{equation}
where $k=1$ for $a=b$ and $k=0$ for $a\neq b$. For small applied voltages
satisfying $|\varepsilon _{{\scriptscriptstyle F}l}-\varepsilon _{{%
\scriptscriptstyle F}r}|\ll \varepsilon _{{\scriptscriptstyle F}r,l}\simeq
\varepsilon _{\scriptscriptstyle F}$, we introduce the tunneling
conductance, ${\rm G}(\Delta )$, through the relation $J_{\bot }={\rm G}%
(\Delta )V$. The interwell voltage, $V$, is connected with the quasi-Fermi
level difference by the relation $V=(\varepsilon _{{\scriptscriptstyle F}%
l}-\varepsilon _{{\scriptscriptstyle F}r})/e$. Then from Eq. (\ref{10}) it
follows that the tunneling conductance can be written as 
\begin{equation}
{\rm G}(\Delta )=\frac{(eT)^{2}}{2\pi \hbar }\frac{2}{L^{2}}\int d{\bf x}%
\int d{\bf x^{\prime }}\sum_{ab={\scriptscriptstyle RA}}(-1)^{k}\left\langle
\left\langle {\cal G}_{l\varepsilon _{\scriptscriptstyle F}}^{a}({\bf x},%
{\bf x^{\prime }}){\cal G}_{r\varepsilon _{\scriptscriptstyle F}}^{b}({\bf %
x^{\prime }},{\bf x})\right\rangle \right\rangle .  \label{11}
\end{equation}
Furthermore according to Eq. (\ref{4}), the short-range potentials in the $l$%
-QW and $r$-QW are statistically independent, then the two-particle
correlation function $\left\langle \left\langle ...\right\rangle
\right\rangle $ in Eq. (\ref{11}) can be rewritten exactly in terms of the
Green's functions $G_{j\varepsilon }^{a}({\bf x},{\bf x^{\prime }})=\langle 
{\cal G}_{j\varepsilon }^{a}({\bf x},{\bf x^{\prime }})\rangle $ averaged
over the short-range potentials. The Dyson equation for this Green's
functions is written as 
\begin{equation}
(\tilde{h}_{j}-\varepsilon )G_{j\varepsilon }^{a}({\bf x},{\bf x^{\prime }}%
)+\int d{\bf x}_{1}\Sigma _{j\varepsilon }^{a}({\bf x},{\bf x}%
_{1})G_{j\varepsilon }^{a}({\bf x}_{1},{\bf x^{\prime }})=\delta ({\bf x}-%
{\bf x^{\prime }}).  \label{12}
\end{equation}
Here the Hamiltonians $\tilde{h}_{l,r}$ coincide with those given in Eq. (%
\ref{3}), without the short-range potentials $\widetilde{U}_{l,r{\bf x}}$,
and $\Sigma _{j\varepsilon }^{a}({\bf x},{\bf x^{\prime }})$ is the
self-energy function. For $\delta $-correlated potentials, we have to use $%
\Sigma _{j\varepsilon }^{a}({\bf x},{\bf x^{\prime }})\propto \delta ({\bf x}%
-{\bf x^{\prime }})$. Neglecting the renormalization of energy spectra, we
rewrite Eq. (\ref{12}) in terms of the broadening energy $\gamma _{j}$ of
the $j$th QW as 
\begin{equation}
(\tilde{h}_{j}-\varepsilon \mp i\gamma _{j})G_{j\varepsilon }^{%
\scriptscriptstyle R,A}({\bf x},{\bf x^{\prime }})=\delta ({\bf x}-{\bf %
x^{\prime }}),  \label{13}
\end{equation}
where the upper sign corresponds to $G^{R}$ and the lower one to $G^{A}$.

It is convenient to write the Green's functions through path integrals as 
\cite{11} 
\begin{eqnarray}
G_{l\varepsilon }^{\scriptscriptstyle R}({\bf x},{\bf x^{\prime }}) &=&\frac{%
i}{\hbar }\int_{-\infty }^{0}dte^{-i(\varepsilon +i\gamma _{l}-\Delta
)t/\hbar }\int_{{\bf x}_{o}={\bf x^{\prime }}}^{{\bf x}_{t}={\bf x}}{\cal D}%
\{{\bf x}_{\tau }\}\exp \left[ -\frac{i}{2\hbar }\int_{0}^{t}d\tau (m\dot{%
{\bf x}}_{\tau }^{2}-\delta U_{{\bf x}_{\tau }})\right] ,  \nonumber \\
G_{r\varepsilon }^{\scriptscriptstyle R}({\bf x},{\bf x^{\prime }}) &=&\frac{%
i}{\hbar }\int_{-\infty }^{0}dte^{-i(\varepsilon +i\gamma _{r})t/\hbar
}\int_{{\bf x}_{o}={\bf x^{\prime }}}^{{\bf x}_{t}={\bf x}}{\cal D}\{{\bf x}%
_{\tau }\}\exp \left[ -\frac{i}{2\hbar }\int_{0}^{t}d\tau (m\dot{{\bf x}}%
_{\tau }^{2}+\delta U_{{\bf x}_{\tau }})\right] \;,  \label{14}
\end{eqnarray}
and $G_{j\varepsilon }^{\scriptscriptstyle A}({\bf x},{\bf x^{\prime }}%
)=G_{j\varepsilon }^{\scriptscriptstyle R}({\bf x^{\prime }},{\bf x})^{\ast
} $. The average over the non-screened large-scale potential in Eq. (\ref{11}%
), for a Gaussian-type random potential $\delta U_{{\bf x}}$, is performed
using the well-known exact formula 
\begin{equation}
\left\langle \exp \left( \int d{\bf x}f_{{\bf x}}\delta U_{{\bf x}}\right)
\right\rangle =\exp \left[ \frac{1}{2}\int d{\bf x}\int d{\bf x^{\prime }}f_{%
{\bf x}}w(|{\bf x}-{\bf x^{\prime }}|)f_{{\bf x^{\prime }}}\right] ,
\label{15}
\end{equation}
for some arbitrary function $f_{{\bf x}}$. Since random potentials are
involved in both path integrals, we choose these functions as 
\[
f_{{\bf x}}=\pm \left( i/2\hbar \right) \int_{0}^{t_{1}}d\tau _{1}\delta (%
{\bf x}-{\bf x}_{\tau _{1}})\pm \left( i/2\hbar \right)
\int_{0}^{t_{2}}d\tau _{2}\delta ({\bf x}-{\bf x}_{\tau _{2}}). 
\]
Using these transformations in the correlation functions of Eq. (\ref{11}),
we finally obtain 
\begin{eqnarray}
\sum_{ab={\scriptscriptstyle RA}}(-1)^{k}\left\langle G_{l\varepsilon _{%
\scriptscriptstyle F}}^{a}({\bf x},{\bf x^{\prime }})G_{r\varepsilon _{%
\scriptscriptstyle F}}^{b}({\bf x^{\prime }},{\bf x})\right\rangle &=&\frac{1%
}{\hbar ^{2}}\int_{-\infty }^{0}dt_{1}e^{(\gamma _{\scriptscriptstyle %
l}+i\Delta )t_{1}/\hbar }\int_{-\infty }^{0}dt_{2}e^{\gamma _{%
\scriptscriptstyle r}t_{2}/\hbar }  \nonumber \\
&&\times \int_{{\bf x}_{o}={\bf x^{\prime }}}^{{\bf x}_{t_{1}}={\bf x}}{\cal %
D}\{{\bf x}_{\tau }\}\left\{ e^{-i\varepsilon _{\scriptscriptstyle %
F}(t_{1}+t_{2})/\hbar }\int_{{\bf y}_{o}={\bf x}}^{{\bf y}_{t_{2}}={\bf %
x^{\prime }}}{\cal D}\{{\bf y}_{\tau }\}\exp \left[ -{\cal S}_{+}(t_{1}t_{2}|%
{\bf x}_{\tau },{\bf y}_{\tau })\right] \right.  \nonumber \\
&&\left. +e^{-i\varepsilon _{\scriptscriptstyle F}(t_{1}-t_{2})/\hbar }\int_{%
{\bf y}_{o}={\bf x^{\prime }}}^{{\bf y}_{t_{2}}={\bf x}}{\cal D}\{{\bf y}%
_{\tau }\}\exp \left[ -{\cal S}_{-}(t_{1}t_{2}|{\bf x}_{\tau },{\bf y}_{\tau
})\right] \right\} +\text{c.c}.~~,  \label{16}
\end{eqnarray}
where the two-particle actions ${\cal S}_{\pm }(t_{1}t_{2}|{\bf x}_{\tau },%
{\bf y}_{\tau })$ are written in the form 
\begin{eqnarray}
{\cal S}_{\pm }(t_{1}t_{2} &|&{\bf x}_{\tau },{\bf y}_{\tau })=\frac{im}{%
2\hbar }\left[ \int_{0}^{t_{1}}d\tau \dot{{\bf x}}_{\tau }^{2}\pm
\int_{0}^{t_{2}}d\tau \dot{{\bf y}}_{\tau }^{2}\right]  \nonumber \\
&&+\frac{1}{8\hbar ^{2}}\int_{0}^{t_{1}}d\tau \int_{0}^{t_{1}}d\tau ^{\prime
}w(|{\bf x}_{\tau }-{\bf x}_{\tau ^{\prime }}|)  \nonumber \\
&&+\frac{1}{8\hbar ^{2}}\int_{0}^{t_{2}}d\tau \int_{0}^{t_{2}}d\tau ^{\prime
}w(|{\bf y}_{\tau }-{\bf y}_{\tau ^{\prime }}|)  \nonumber \\
&&\mp \frac{1}{4\hbar ^{2}}\int_{0}^{t_{1}}d\tau \int_{0}^{t_{2}}d\tau
^{\prime }w(|{\bf x}_{\tau }-{\bf y}_{\tau ^{\prime }}|).  \label{17}
\end{eqnarray}

Substituting Eq. (\ref{16}) into Eq. (\ref{11}) and making convenient change
of variables (in particular, separating the straight path according to ${\bf %
x}_{\tau }\rightarrow \lbrack {\bf u}\tau /t_{1}+{\bf x}_{\tau }]$ and ${\bf %
y}_{\tau }\rightarrow \lbrack {\bf u}(t_{2}-\tau )/t_{2}+{\bf y}_{\tau }]$,
for integral from $\exp (-{\cal S}_{+})$, or ${\bf y}_{\tau }\rightarrow
\lbrack {\bf u}\tau /t_{2}+{\bf y}_{\tau }]$, for integral from $\exp (-%
{\cal S}_{-})$), we can express ${\rm G}(\Delta )$ in terms of contour
integrals as

\begin{eqnarray}
{\rm G}(\Delta ) &=&\frac{(eT)^{2}}{\pi \hbar ^{3}}\int d{\bf u}%
\int_{-\infty }^{0}dt_{1}e^{(\gamma _{\scriptscriptstyle l}+i\Delta
)t_{1}/\hbar }\int_{-\infty }^{0}dt_{2}e^{\gamma _{\scriptscriptstyle %
r}t_{2}/\hbar }  \nonumber \\
&&\times \oint {\cal D}\{{\bf x}_{\tau }\}\oint {\cal D}\{{\bf y}_{\tau
}\}\sum_{\pm }\left\{ e^{-i\varepsilon _{F}(t_{1}\pm t_{2})/\hbar }\exp %
\left[ -\frac{im}{2\hbar }{\bf u}^{2}(t_{1}^{-1}\pm t_{2}^{-1})\right]
\right.  \nonumber \\
&&\left. \times \exp \left[ -\frac{im}{2\hbar }\left( \int_{0}^{t_{1}}d\tau 
\dot{{\bf x}}_{\tau }^{2}\pm \int_{0}^{t_{2}}d\tau \dot{{\bf y}}_{\tau
}^{2}\right) \right] \exp \left[ -K_{\pm }(t_{1},t_{2},{\bf x}_{\tau },{\bf y%
}_{\tau })\right] 2\right\} +\text{c.c}.\;,  \label{18}
\end{eqnarray}
where ${\bf u}={\bf x}-{\bf x^{\prime }}$. The contributions of non-screened
potentials to the correlation function is given by the factors 
\begin{eqnarray}
K_{\pm }(t_{1},t_{2},{\bf x}_{\tau },{\bf y}_{\tau }) &=&\frac{1}{8\hbar ^{2}%
}\int_{0}^{t_{1}}d\tau \int_{0}^{t_{1}}d\tau ^{\prime }w(|{\bf x}_{\tau }-%
{\bf x}_{\tau ^{\prime }}+{\bf u}(\tau -\tau ^{\prime })/t_{1}|)  \nonumber
\\
&&+\frac{1}{8\hbar ^{2}}\int_{0}^{t_{2}}d\tau \int_{0}^{t_{2}}d\tau ^{\prime
}w(|{\bf y}_{\tau }-{\bf y}_{\tau ^{\prime }}\pm {\bf u}(\tau -\tau ^{\prime
})/t_{2}|)  \nonumber \\
&&\mp \frac{1}{4\hbar ^{2}}\int_{0}^{t_{1}}d\tau \int_{0}^{t_{2}}d\tau
^{\prime }w_{\pm }(|{\bf x}_{\tau }-{\bf y}_{\tau ^{\prime }}+{\bf u(}\tau
/t_{1}\pm \tau ^{\prime }/t_{2})|)\;.  \label{19}
\end{eqnarray}
with $w_{-}(|{\bf z}|)=w(|{\bf z}|)$ and $w_{+}(|{\bf z}|)=w(|{\bf z-u}|)$.
Note that $K_{+}$ comes from averaging both retarded or both advanced
Green's functions while $K_{-}$\ corresponds to averaging the product of
retarded and advanced Green's functions.

\section{Lineshape of the conductance peak}

In order to calculate the path integrals in Eq. (\ref{18}), we will neglect
in Eqs. (\ref{18}), (\ref{19}) deviations ${\bf x}_{\tau }\;$and ${\bf y}%
_{\tau }$ in the arguments of the correlation function $w(|...|)$ \ by
supposing that these deviations are smaller than $\ell _{c}$, i.e., using
the approach of straightforward trajectory in Eq. (\ref{19}). We justify
such an approximation in Appendix B, where the optimal fluctuation method 
\cite{13} is used, in order to extract the optimal trajectories which give
the maximal contribution to the path integrals. With this approximation, we
can calculate the path integrals for the free motion exactly and the
conductance, given by Eq. (\ref{18}), is rewritten as

\begin{eqnarray}
{\rm G}(\Delta ) &=&\frac{(eT)^{2}}{\pi \hbar ^{3}}\int d{\bf u}%
\int_{-\infty }^{0}dt_{1}e^{(\gamma _{\scriptscriptstyle l}+i\Delta
)t_{1}/\hbar }\int_{-\infty }^{0}dt_{2}e^{\gamma _{\scriptscriptstyle %
r}t_{2}/\hbar }\frac{(m/2\pi \hbar )^{2}}{t_{1}t_{2}}  \nonumber \\
&&\times \sum_{\pm }\left[ \mp e^{-i\varepsilon _{F}(t_{1}\pm t_{2})/\hbar
}\exp \left( -\frac{imu^{2}}{2\hbar }(t_{1}^{-1}\pm t_{2}^{-1})-K_{\pm
}(t_{1},t_{2},u)\right) \right] +\text{c.c}.\;,  \label{20}
\end{eqnarray}
where the factors $K_{\pm }$ are reduced to 
\begin{eqnarray}
K_{\pm }(t_{1},t_{2},u) &=&\frac{1}{8\hbar ^{2}}\int_{0}^{t_{1}}d\tau
\int_{0}^{t_{1}}d\tau ^{\prime }w(u|\tau -\tau ^{\prime }|/t_{1})  \nonumber
\\
&&+\frac{1}{8\hbar ^{2}}\int_{0}^{t_{2}}d\tau \int_{0}^{t_{2}}d\tau ^{\prime
}w(u|\tau -\tau ^{\prime }|/t_{2})  \nonumber \\
&&\mp \frac{1}{4\hbar ^{2}}\int_{0}^{t_{1}}d\tau \int_{0}^{t_{2}}d\tau
^{\prime }w_{\pm }(|{\bf u(}\tau /t_{1}\pm \tau ^{\prime }/t_{2})|)\;.
\label{21}
\end{eqnarray}
Let us for a moment ignore the contribution from the terms with the upper
sign in Eq. (\ref{20}). Defining new variables $x=\tau /t_{1,2}$ and $%
x^{\prime }=\tau ^{\prime }/t_{1,2}$ in the factor $K_{-}(t_{1},t_{2},u)$,
we obtain the conductance in the form 
\begin{eqnarray}
{\rm G}(\Delta ) &=&\frac{(eT)^{2}}{\pi \hbar ^{3}}\int d{\bf u}%
\int_{-\infty }^{0}dt_{1}e^{(\gamma _{\scriptscriptstyle l}+i\Delta
)t_{1}/\hbar }\int_{-\infty }^{0}dt_{2}e^{\gamma _{\scriptscriptstyle %
r}t_{2}/\hbar }\frac{(m/2\pi \hbar )^{2}}{t_{1}t_{2}}  \nonumber \\
&&\times e^{-i\varepsilon _{F}(t_{1}-t_{2})/\hbar }\exp \left[ -\frac{imu^{2}%
}{2\hbar }\left( t_{1}^{-1}-t_{2}^{-1}\right) -\frac{(t_{1}+t_{2})^{2}}{%
8\hbar ^{2}}W\left( \frac{u}{\ell _{c}}\right) \right] +\text{c.c}.\;,
\label{23}
\end{eqnarray}
where the large-scale correlation function is transformed as $W(u/\ell _{c})=%
\overline{\delta \varepsilon }^{2}\int_{0}^{1}dx\int_{0}^{1}dx^{\prime }\exp
[-(u/l_{c})^{2}(x-x^{\prime })^{2}]$ and can be rewritten as 
\begin{equation}
W(x)/\overline{\delta \varepsilon }^{2}=\sqrt{\pi }x^{-1}%
\mathop{\rm erf}%
(x)-x^{-2}[1-e^{-x^{2}}],  \label{24}
\end{equation}
and $%
\mathop{\rm erf}%
(x)$ is the error function. Introducing new time variables $\tau
=t_{1}-t_{2} $ and $t=(t_{1}+t_{2})/2$ it follows that 
\begin{eqnarray}
{\rm G}(\Delta ) &=&\frac{(eT)^{2}}{\pi \hbar ^{3}}\int d{\bf u}%
\int_{-\infty }^{0}dt\int_{2t}^{-2t}d\tau e^{(\gamma t+\Delta \gamma \tau
)/\hbar }\frac{(m/2\pi \hbar )^{2}}{t^{2}-\tau ^{2}/4}  \nonumber \\
&&\times e^{i[\Delta (t+\tau /2)-\varepsilon _{F}\tau ]/\hbar }\exp \left[ 
\frac{imu^{2}}{2\hbar }\frac{\tau }{t^{2}-\tau ^{2}/4}-\frac{t^{2}}{2\hbar
^{2}}W\left( \frac{u}{\ell _{c}}\right) \right] +\text{c.c.}\;,  \label{25}
\end{eqnarray}
where $\gamma =\gamma _{l}+\gamma _{r}$ and $\Delta \gamma =(\gamma
_{l}-\gamma _{r})/2$ are the total collision-induced broadening and the
broadening difference in $l$- and $r$-QWs respectively. Since the time scale
of $\tau $ is of the order of $\hbar /\varepsilon _{\scriptscriptstyle F}$
and a typical $t$ is of the order of $\hbar /\gamma _{\scriptscriptstyle %
eff} $ in the integrals of Eq. (\ref{25}), we can replace $t^{2}-\tau ^{2}/4$
by $t^{2}$, due to the quasiclassical condition $\gamma _{\scriptscriptstyle %
eff}\ll \varepsilon _{\scriptscriptstyle F}$ and the integration over $\tau $
gives us $2\pi \hbar \delta \lbrack \varepsilon _{\scriptscriptstyle %
F}-m(u/t)^{2}/2]$. After straightforward integration over ${\bf u}$ we
finally obtain 
\begin{equation}
{\rm G}(\Delta )\simeq \left( \frac{eT}{\hbar }\right) ^{2}\rho _{%
\scriptscriptstyle2D}\int_{-\infty }^{0}dte^{(\gamma +i\Delta )t/\hbar }\exp %
\left[ -\frac{t^{2}}{2\hbar ^{2}}W\left( \frac{v_{\scriptscriptstyle F}t}{%
\ell _{c}}\right) \right] +\text{c.c}.\;,  \label{26}
\end{equation}
where $\rho _{\scriptscriptstyle2D}=m/\pi \hbar ^{2}$ is the 2D density of
states, the correlation function is given by Eq. (\ref{24}), and $v_{%
\scriptscriptstyle F}$ is the Fermi velocity.

Consider first the limiting case of the local response, assuming 
\begin{equation}
(v_{\scriptscriptstyle F}\hbar /\gamma _{\scriptscriptstyle eff}\ell
_{c})^{2}\ll 1.  \label{27}
\end{equation}
where the effective HWHM due to both contribution from collision processes
and inhomogeneous broadening is determined by ${\rm G}(\gamma _{%
\scriptscriptstyle eff})={\rm G}(0)/2$. Under such a condition the
correlation function (\ref{24}) assumes the form $W(u/\ell _{c})\approx W(0)=%
\overline{\delta \varepsilon }^{2}$ and for conductance lineshape from Eq. (%
\ref{26}) it follows that \cite{14} 
\begin{eqnarray}
{\rm G}(\Delta ) &=&2\left( \frac{eT}{\hbar }\right) ^{2}\rho _{%
\scriptscriptstyle2D}\int_{-\infty }^{0}dte^{\gamma t/\hbar -(\overline{%
\delta \varepsilon }t/\sqrt{2}\hbar )^{2}}\cos [(\Delta /\hbar )t]  \nonumber
\\
&=&\frac{(eT)^{2}}{\hbar }\rho _{\scriptscriptstyle2D}\left\{ 
\begin{array}{ll}
2\gamma /(\Delta ^{2}+\gamma ^{2})\;, & \overline{\delta \varepsilon }\ll
\gamma \\ 
\left( \sqrt{2\pi }/\overline{\delta \varepsilon }\right) \exp \left[
-(\Delta /\sqrt{2}\;\overline{\delta \varepsilon })^{2}\right] & \overline{%
\delta \varepsilon }\gg \gamma
\end{array}
\right. ,  \label{28}
\end{eqnarray}
where the limiting cases determine the Lorentzian or Gaussian lineshape.
Notice that the Lorentzian shape of the peak tails is always found for big
enough $|\Delta |$.

Now, consider the variables $t_{1,2}$ in the integral with the upper sign in
Eq. (\ref{20}). They are estimated of the order of $\hbar /\varepsilon _{%
\scriptscriptstyle F}.$ Thus $u$ is of the order of $\hbar /\sqrt{%
m\varepsilon _{\scriptscriptstyle F}}\ll \ell _{c}$, and as a result, we can
replace the factor $\exp [-K_{+}(t_{1},t_{2},u)]$ by the expression $\exp
[-(t_{1}-t_{2})^{2}w(0)/(8\hbar ^{2})]$ where the exponential factor is of
the order of $(\overline{\delta \varepsilon }/\varepsilon _{%
\scriptscriptstyle F})^{2}\ll 1$ and the expression can be approximated by
the unity. After straightforward integrations over ${\bf u}$ and $\tau $, we
are left with an integral over $t$ given by

\begin{equation}
-i\frac{(eT)^{2}\rho _{2D}}{\pi \hbar ^{2}}\int_{-\infty }^{0}dte^{(\gamma
_{l}+\gamma _{r}+i\Delta )t/\hbar }e^{-i2\varepsilon _{F}t/\hbar }+\text{c.c}%
.\approx \frac{(eT)^{2}\rho _{2D}}{2\pi \hbar \varepsilon _{%
\scriptscriptstyle F}}.  \label{22}
\end{equation}
Such a contribution can be discarded in comparison with the results from
Eqs. (\ref{26}) and (\ref{28}) because this term leads to corrections of the
order of $\gamma _{\scriptscriptstyle eff}/\varepsilon _{\scriptscriptstyle %
F}$.

In Fig. 2, we plot ${\rm G}(\Delta )$, using Eq. (\ref{28}), as function of $%
\Delta /\gamma $ for different relative contributions of the scattering
processes and of the inhomogeneous broadening, i.e., for different ratios $%
\overline{\delta \varepsilon }/\gamma $. The solid, dashed, dotted,
dot-dashed, and dot-dot-dashed curves in Fig. 2 correspond to $\overline{%
\delta \varepsilon }/\gamma =0.3,$ $0.6,$ $1,$ $3$, and $6$, respectively.
The factor ${\rm G}_{L}(0)=2(eT)^{2}\rho _{2D}/\hbar \gamma $ is obtained by
after putting $\Delta =0$ and $\overline{\delta \varepsilon }=0$ in Eq. (\ref
{28}). We see in Fig. 2 that the change from a Lorentzian and a Gaussian
lineshapes depends also on the dimensionless ratio $|\Delta |/\gamma $.

If we take the opposite limit to the inequality (\ref{27}), i.e., we are
dealing now with relatively short $\ell _{c}$, then $W(v_{\scriptscriptstyle %
F}t/\ell _{c})$ in Eq. (\ref{26}) has to be approximated by $W(x)/\overline{%
\delta \varepsilon }^{2}\approx \sqrt{\pi }x^{-1}$. As a result, the
Lorentzian lineshape is obtained, from Eq. (\ref{26}), as

\begin{equation}
{\rm G}(\Delta )=\frac{(eT)^{2}}{\hbar }\rho _{\scriptscriptstyle2D}\frac{2%
\overline{\gamma }_{eff}}{\Delta ^{2}+\overline{\gamma }_{eff}^{2}}
\label{29}
\end{equation}
where the effective HWHM is now defined by $\overline{\gamma }_{%
\scriptscriptstyle eff}=\gamma \lbrack 1+(\sqrt{\pi }/2)(\overline{\delta
\varepsilon }^{2}\ell _{c}/\hbar v_{F}\gamma )]$. So, by increasing $\ell
_{c}$, a transition from the Lorentzian to a Gaussian lineshape is obtained
for $\overline{\delta \varepsilon }>\gamma $.

This peak modification is illustrated in Fig. 3, where ${\rm G}(\Delta )$,
calculated from Eq. (\ref{26}), is displayed for $\overline{\delta
\varepsilon }/\gamma =4.6$. In Fig. 3, the solid, dashed, dotted, and
dot-dashed curves correspond to $\hbar v_{F}/\ell _{c}\gamma =15,$ $3.5,$ $%
0.7$, and $0.2$, respectively. Notice that the solid curve corresponds very
closely to the Lorentzian given by Eq. (\ref{29}), while the dotted and the
dot-dashed curves are practically coincident Gaussians. In Fig. 4, we plot $%
\gamma _{eff}/\gamma $, calculated from Eq. (\ref{26}), as a function of $%
\hbar v_{F}/\ell _{c}\gamma $ for decreasing values of $\overline{\delta
\varepsilon }/\gamma =4.6$ (top), $2.3,$ $1.5,$ $1.1,$ $0.8$, and $0.3$
(bottom). We point out that all curves give $\gamma _{\scriptscriptstyle %
eff} $ in the local regime for $\hbar v_{F}/\ell _{c}\gamma =0$. Thus, from
Fig. 4, it is seen that nonlocal effects essentially make $\gamma _{%
\scriptscriptstyle eff}$ decrease for $\overline{\delta \varepsilon }/\gamma %
\agt 1$.

Now we apply the present model calculation to interpret the experimental
data of Ref. \cite{5}. We will consider the results in the low-temperature
regime (less than $2$ K), where the measured HWHM $\gamma _{%
\scriptscriptstyle eff}$ is practically independent of the temperature. We
assume that only the inverted heterointerface for each QW has essential
roughness \cite{15} due to one-monolayer variations ($\overline{a}\approx 2.5
$\AA ) and, according to Appendix A, the characteristic energy is estimated
as $\overline{\delta \varepsilon }\approx 0.46$ meV. The hard-wall model for
a QW is used here to calculate $\overline{\varepsilon }$ from (\ref{(A.6)}).
For sample A, with electron density $1.6\times 10^{11}$ cm$^{-2}$, we assume
that $\gamma \approx 0.1$ meV (from mobility data), $\ell _{c}\approx 700$ 
\AA\ and obtain from the pertinent curve, for $\overline{\delta \varepsilon }%
/\gamma =4.6$, in Fig. 4, that $\gamma _{\scriptscriptstyle eff}\approx 0.22$
meV, which coincides with the experimental data and corresponds to $\hbar
v_{F}/\ell _{c}\gamma =15$ (the solid square in Fig. 4). As a consequence,
the Lorentzian lineshape given by the solid curve in Fig. 3, is appropriate
for sample A. Furthermore, for sample B, with density $1.5\times 10^{11}$ cm$%
^{-2}$, by assuming that $\gamma \approx 0.2$ meV and $\ell _{c}\approx 1400$
\AA , we obtain $\gamma _{\scriptscriptstyle eff}\approx 0.45$ meV from the
pertinent curve, for $\overline{\delta \varepsilon }/\gamma =2.3$, in Fig.
4, after using the calculated value $\hbar v_{F}/\ell _{c}\gamma =3.8$ (the
solid triangle). This value is in good agreement with the experimental
result of Ref. \cite{5}. For sample C, with density $0.8\times 10^{11}$ cm$%
^{-2}$ and assuming $\gamma \approx 0.2$ meV and $\ell _{c}\approx 1000$ \AA %
, we have $\gamma _{\scriptscriptstyle eff}\approx 0.45$ meV (indicated by
the solid triangle in Fig. 4), i.e., the same as for sample B and also
coincident with experimental observations \cite{5}. Then a good agreement
with the experimental results is found for the case of single-side
variations of heterointerfaces (see Ref. \cite{16} about the case of
two-side variations).

Notice that the change from a local regime of tunneling (for long-range
fluctuations of QW widths) to the general nonlocal case may be found by
varying the temperature, that controls the relative contributions of
homogeneous and inhomogeneous broadening. In Fig. 5 we plot the lineshapes $%
{\rm G}(\Delta )$, for temperatures $\Theta $ in the range 0.7 K - 10 K,
calculated from Eq. (\ref{26}), for sample B parameters taken from Ref. \cite
{5}. Now we have to add the thermal {\it e - e} scattering contribution to $%
\gamma $, leading to a renormalized value $\gamma (\Theta )$, which we
approximate in the same way as in Ref. \cite{5} (see the solid curve in Fig.
3 of Ref. \cite{5}). In Fig. 5, the solid, dashed, dotted and dot-dashed
curves correspond to $\Theta =0.7$, $5$, $7$, and $10$ K, respectively. One
can see that by decreasing the temperature, the linewidth becomes smaller
and the shape of peak is changed. While for $\Theta =0.7$ K, we observe a
Lorentzian form of the peak due to strong nonlocal effects, manifested by
inhomogeneous broadening induced by non-screened large-scale fluctuations,
nonlocal effects are quite weak for $\Theta =10$ K. At this temperature the
local regime prevails and the lineshape is the interplay of Gaussian and
Lorentzian forms, given by Eq. (\ref{28}), because in this case $\overline{%
\delta \varepsilon }/\gamma (\Theta )\approx 0.5$ and $\hbar v_{F}/\ell
_{c}\gamma (\Theta )\approx 0.8$ the peak behavior is slightly more
Lorentzian than Gaussian.

\section{Concluding remarks}

In the present work, we have introduced a new electron scattering mechanism
by long-scale non-screened roughness of heterointerfaces which contributes
to the inhomogeneous broadening of the tunneling conductance peak in coupled
DQWs. We have done a systematic analysis of this peak shape by taking into
account the interplay between the introduced mechanism, and the usual
homogeneous scattering broadening. A detailed comparison of the HWHM of the
peak with experimental results revealed that the considered mechanism is
relevant to interpret the data of Ref. \cite{5}, because in the experimental
conditions strong nonlocal effects are manifested, through the proposed
mechanism of inhomogeneous broadening, which modify drastically the
lineshape (from the Gaussian to a Lorentzian) and the HWHM of the peaks. We
call the attention to general considerations for the mechanism of appearance
of non-screened long wavelength variations of the scattering potential,
given in Appendix A, which should be relevant not only for the problem of
inter-QW tunneling current, addressed here, but also for the study of
general transport and optical properties of doped DQWs.

Let us discuss the approximations used in our treatment. The single-electron
approximation for the tunneling Hamiltonian in Sec. II is a generally
accepted model and the expression for the tunneling current in the
homogeneous case corresponds to the Bardeen's approach.\cite{17} Here smooth
variations of boundaries lead to changes of the levels of $l-$ and $r-$QWs
in the tunneling Hamiltonian, and we assume that small modifications of the
tunneling matrix element can be neglected. For a discussion of the latter
approximation, see Ref. \cite{9}. The approximation for considering the
screening ``on average'' in the introduced large-scale potential consists in
supposing that the correlation length $\ell _{c}$ is large in comparison
with the Bohr radius and with transverse dimensions of the DQW structure as
well. Since ${\rm G}(\Delta )$ depends on the sum of the scattering
broadening of different levels, we believe that the introduction of
phenomenological parameters $\gamma _{l,r}$ instead of a detailed
consideration the self-energy functions, does not lead to the omission of
important contributions. We have also used the quasi-classical description
for longitudinal motion which is valid when $\varepsilon _{%
\scriptscriptstyle
F}\gg \gamma _{\scriptscriptstyle eff}$ in the calculation of the path
integrals in Eq. (\ref{18}) and for integration over ${\bf u}$ and $\;\tau $
in Eq. (\ref{25}). We have assumed that variations of potential are
sufficiently weak such that the acceleration of an electron due to
long-scale random force (quasi-electric field) on a length of the order of $%
\ell _{c}$ is insignificant. We have considered that the inverted
AlGaAs-GaAs heterointerface is much more rougher than the normal GaAs-AlGaAs
heterointerface based on the results of Ref. \cite{15} for QWs similar to
those used in DQW structures of Ref. \cite{5}.

To conclude, we now discuss the possibility of a more reliable test of the
described mechanism of inhomogeneous broadening. In further experimental
studies of the tunneling conductance it is necessary to make a more detailed
analysis of the lineshape transition (from the Lorentzian to Gaussian form),
and comparison between HWHM data measured in samples grown in different
conditions. Because the considered mechanism modifies also in-plane
transport coefficients and optical properties of DQWs with long-wavelength
inhomogeneities, further measurements as well theoretical studies of these
phenomena are necessary.

We believe that the present work establishes an essential contribution of
large-scale non-screened fluctuations to the broadening of the tunneling
conductance peak in DQWs in agreement with experimental results.\cite{5}

\acknowledgements

This work was supported by grants Nos. 95/0789-3 and 98/10192-2 from Funda\c{%
c}\~{a}o de Amparo \`{a} Pesquisa de S\~{a}o Paulo (FAPESP). O. G. B. and N.
S. are grateful to Conselho Nacional de Desenvolvimento Cient\'{i}fico e
Tecnol\'{o}gico (CNPq) for research fellowships.\newline

\ \newline
\appendix

\section{Non-screened variations of the random potential}

Below we evaluate the non-screened random contributions to the potential $%
\pm \delta U_{{\bf x}}$ that appears in Eqs. (\ref{2}) - (\ref{4}). The 2D
Fourier transform of the screening potential, $V_{{\bf q}z}$, is determined
by the Poisson equation 
\begin{equation}
\left( \frac{d^{2}}{dz^{2}}-q^{2}\right) V_{{\bf q}z}=-\frac{4\pi e^{2}}{%
\epsilon }\delta n_{{\bf q}z},  \label{A.1}
\end{equation}
where $\delta n_{{\bf q}z}$ is the concentration induced by the total
large-scale potential. Neglecting the overlap of $l$- and $r$- orbitals $%
\varphi _{jz}$ ($j=l,r$) we use in Eq. (\ref{A.1}) the expansion 
\begin{equation}
\delta n_{{\bf q}z}=\sum_{j=l,r}\delta n_{{\bf q}j}\varphi _{jz}^{2},
\label{A.2}
\end{equation}
where $\delta n_{{\bf q}j}=-\rho _{\scriptscriptstyle2D}(\overline{U}_{{\bf q%
}j}+V_{{\bf q}j})$ is the in-plane induced concentration in the $j$th QW due
to slow variations of the potential while the non-diagonal components of $%
\delta \hat{n}_{{\bf x}}$ are small due to weak overlap of $l$- and $r$%
-orbitals. Note that, for the general case of slowly varying
heterointerfaces, $\varphi _{j{\bf x}z}$ should depend on the in-plane
coordinate ${\bf x}$ but for the 2D case ($\varepsilon _{\scriptscriptstyle %
F}\ll \varepsilon _{j}$) such in-plane variations of orbitals are not
important

The solution of Eq.(\ref{A.1}) assumes the form 
\begin{equation}
V_{{\bf q}z}=\frac{2\pi e^{2}}{\epsilon q}\int dz^{\prime }e^{-q|z-z^{\prime
}|}\delta n_{{\bf q}z^{\prime }},  \label{A.3}
\end{equation}
where for the case $q\alt\ell _{c}^{-1}\ll d^{-1}$ we have $\exp
(-q|z-z^{\prime }|)\simeq 1$. Then the diagonal components $V_{{\bf q}j}$, $%
j=l,r$, are expressed in terms of the total concentration as 
\begin{equation}
V_{{\bf q}j}=\int dz\varphi _{jz}^{2}V_{{\bf q}z}=\frac{2\pi e^{2}}{\epsilon
q}\sum_{s=l,r}\delta n_{{\bf q}s},  \label{A.4}
\end{equation}
Substituting $\delta n_{{\bf q}j}$\ in Eq. (\ref{A.4}), we have 
\begin{equation}
V_{{\bf q}j}=-\frac{4}{qa_{\scriptscriptstyle B}}\sum_{s=l,r}(\overline{U}_{%
{\bf q}s}+V_{{\bf q}s}),  \label{A.5}
\end{equation}
where $a_{\scriptscriptstyle B}$ is the Bohr radius. Since the right-hand
side of Eq. (\ref{A.5}) does not depend on $j$ we obtain $V_{{\bf q}l,r}=V_{%
{\bf q}}$ and we deal with an averaged screened potential $V_{{\bf q}}$
across the DQWs. If we assume large-scale variations, in particular $\ell
_{c}\gg a_{\scriptscriptstyle B}$, and the condition $qa_{\scriptscriptstyle %
B}/8\ll 1$, we derive the Eq. (\ref{2}), $V_{{\bf q}}\simeq -(\sum_{j}%
\overline{U}_{{\bf q}j})/2$.

Now we present explicit expressions for the large-scale addends to the
matrix Hamiltonian (\ref{1}) due to the random variations of the DQWs
heterointerfaces. Statistically independent boundaries variations are
described by random functions $\delta _{{\bf x}}^{<}$ and $\delta _{{\bf x}%
}^{>}$ (see Fig. 1$b$) so that large-scale potentials take the form $%
\overline{U}_{l,r{\bf x}}\simeq 2\varepsilon _{l,r}(\delta _{{\bf x}%
}^{<}-\delta _{{\bf x}}^{>})/d_{l,r}$, where $\varepsilon _{l,r}$ are the
energies of levels in $l,$ and $r$-QWs with width $d_{l,r}$. As it was shown
in Ref. \cite{8}, the weak contributions due to variations of the tunneling
matrix element may be neglected in the evaluation of Eq. (\ref{8}) up to
second-order in $T$. Substituting $\overline{U}_{l,r{\bf x}}$ into the
correlation functions in Eq. (\ref{4}) and supposing that all interfaces are
statistically equivalents, we obtain the correlation functions as 
\begin{equation}
\overline{W}_{l,r}(|{\bf x}-{\bf x^{\prime }}|)=2\left( \frac{2\overline{%
\varepsilon } \; \overline{a}}{\overline{d}}\right) ^{2}\exp \left[ -\left( 
\frac{{\bf x}-{\bf x^{\prime }}}{\ell _{c}}\right) ^{2}\right] ,
\label{(A.6)}
\end{equation}
where $\overline{\varepsilon }\simeq \varepsilon _{l,r}$, $\overline{d}%
\simeq d_{l,r}$, and $\overline{a}$ is the averaged deviation of
heterointerfaces and $\ell _{c}$ is the correlation length. Finally, the
correlation function $w(|...|)$ in Eq. (\ref{4}) takes the form $w(|{\bf x}-%
{\bf x^{\prime }}|)=\left( \overline{\delta \varepsilon }\right) ^{2}\exp
[-\left( {\bf x}-{\bf x^{\prime }}\right) ^{2}/\ell _{c}^{2}]$ with a
characteristic energy $\overline{\delta \varepsilon }=4\overline{\varepsilon 
}\;\overline{a}/\overline{d}$ for the case of two-side variations. However, $%
\overline{\delta \varepsilon }=2\sqrt{2} \overline{\varepsilon } \; 
\overline{a}/\overline{d}$ for the one-side variation case in a QW when $%
\delta _{{\bf x}}^{<}$ or $\delta _{{\bf x}}^{>}$ is equal to zero and one
of the two correlation functions given by Eq. (\ref{(A.6)}) is set to zero.

\section{Optimal fluctuation method}

The evaluation of the Eq. (\ref{20}) is based on the separation of the
optimal trajectory (with maximal contribution to the path integral) and on
the comparison of typical variations to such a trajectory, $\delta {\bf x}%
_{\tau }$ and $\delta {\bf y}_{\tau }$, with $\ell _{c}$. First variations
of the actions in the exponential factors of Eqs. (\ref{18}), (\ref{19})
with respect to path variations $\delta {\bf x}_{\tau }$ and $\delta {\bf y}%
_{\tau }$ are written as follows 
\begin{eqnarray}
&-&\frac{i}{\hbar }\int_{0}^{t_{1}}d\tau \delta {\bf x}_{\tau }\cdot \{m%
\ddot{{\bf x}}_{\tau }-\frac{i}{2\hbar \ell _{c}^{2}}\int_{0}^{t_{1}}d\tau
^{\prime }\left[ {\bf u}(\tau -\tau ^{\prime })/t_{1}+{\bf x}_{\tau }-{\bf x}%
_{\tau ^{\prime }}\right] w(|{\bf u}(\tau -\tau ^{\prime })/t_{1}+{\bf x}%
_{\tau }-{\bf x}_{\tau ^{\prime }}|)  \nonumber \\
&&\pm \frac{i}{2\hbar \ell _{c}^{2}}\int_{0}^{t_{1}}d\tau ^{\prime }\;\left[ 
{\bf u}\left( \frac{\tau }{t_{1}}-\frac{\tau ^{\prime }}{t_{2}}\right) +{\bf %
x}_{\tau }-{\bf y}_{\tau ^{\prime }}-\frac{{\bf u}}{2}\left( 1\pm 1\right) %
\right] w(|{\bf u}\tau /t_{1}+{\bf x}_{\tau }-{\bf u}\tau ^{\prime }/t_{2}-%
{\bf y}_{\tau ^{\prime }}-\frac{{\bf u}}{2}\left( 1\pm 1\right) |)\},
\label{B.1}
\end{eqnarray}
and 
\begin{eqnarray}
&-&\frac{i}{\hbar }\int_{0}^{t_{2}}d\tau \delta {\bf y}_{\tau }\cdot \{\pm m%
\ddot{{\bf y}}_{\tau }-\frac{i}{2\hbar \ell _{c}^{2}}\int_{0}^{t_{2}}d\tau
^{\prime }[\pm {\bf u}(\tau -\tau ^{\prime })/t_{2}+{\bf y}_{\tau }-{\bf y}%
_{\tau ^{\prime }}]w(|\pm {\bf u}(\tau -\tau ^{\prime })/t_{2}+{\bf y}_{\tau
}-{\bf y}_{\tau ^{\prime }}|)  \nonumber \\
&&\pm \frac{i}{2\hbar \ell _{c}^{2}}\int_{0}^{t_{1}}d\tau ^{\prime }\left[ 
{\bf u}\left( \frac{\tau }{t_{2}}-\frac{\tau ^{\prime }}{t_{1}}\right) -{\bf %
x}_{\tau ^{\prime }}+{\bf y}_{\tau }-\frac{{\bf u}}{2}\left( 1\pm 1\right) %
\right] w(|{\bf u}\tau ^{\prime }/t_{1}+{\bf x}_{\tau ^{\prime }}-{\bf u}%
\tau /t_{2}-{\bf y}_{\tau }+\frac{{\bf u}}{2}\left( 1\pm 1\right) |)\},
\label{B.2}
\end{eqnarray}
where the upper (lower) signs correspond to upper (lower) signs in Eqs. (\ref
{18}), (\ref{19}) and we have transformed correlation functions as follows 
\begin{equation}
w(|{\bf x}+\delta {\bf x}|)-w(|{\bf x}|)\simeq -2{\bf x}\cdot \delta {\bf x}%
\;w(x)/\ell _{c}^{2}.  \label{B.3}
\end{equation}
Since $\delta {\bf x}_{\tau }$ and $\delta {\bf y}_{\tau }$ are arbitrary
variations, the optimal trajectories are determined by the system of
Euler-Lagrange equations as 
\begin{eqnarray}
m\ddot{{\bf x}}_{\tau } &=&\frac{i}{2\hbar \ell _{c}^{2}}\int_{0}^{t_{1}}d%
\tau ^{\prime }[{\bf u}(\tau -\tau ^{\prime })/t_{1}+{\bf x}_{\tau }-{\bf x}%
_{\tau ^{\prime }}]w(|{\bf u}(\tau -\tau ^{\prime })/t_{1}+{\bf x}_{\tau }-%
{\bf x}_{\tau ^{\prime }}|)  \nonumber \\
&&\mp \frac{i}{2\hbar \ell _{c}^{2}}\int_{0}^{t_{2}}d\tau ^{\prime }[{\bf u}%
(\tau /t_{1}-\tau ^{\prime }/t_{2})+{\bf x}_{\tau }-{\bf y}_{\tau ^{\prime
}}-{\bf u}\left( 1\pm 1\right) /2]w(|{\bf u}\tau /t_{1}+{\bf x}_{\tau }-{\bf %
u}\tau ^{\prime }/t_{2}-{\bf y}_{\tau ^{\prime }}-{\bf u}\left( 1\pm
1\right) /2|),  \label{B.4}
\end{eqnarray}
and 
\begin{eqnarray}
\pm m\ddot{{\bf y}}_{\tau } &=&\frac{i}{2\hbar \ell _{c}^{2}}%
\int_{0}^{t_{2}}d\tau ^{\prime }[\pm {\bf u}(\tau -\tau ^{\prime })/t_{2}+%
{\bf y}_{\tau }-{\bf y}_{\tau ^{\prime }}]w(|\pm {\bf u}(\tau -\tau ^{\prime
})/t_{2}+{\bf y}_{\tau }-{\bf y}_{\tau ^{\prime }}|)  \nonumber \\
&&\mp \frac{i}{2\hbar \ell _{c}^{2}}\int_{0}^{t_{1}}d\tau ^{\prime }[{\bf u}%
(\tau /t_{2}-\tau ^{\prime }/t_{1})-{\bf x}_{\tau ^{\prime }}+{\bf y}_{\tau
}-{\bf u}\left( 1\pm 1\right) /2]w(|{\bf u}\tau /t_{2}+{\bf y}_{\tau }-{\bf u%
}\tau ^{\prime }/t_{1}-{\bf x}_{\tau ^{\prime }}-{\bf u}\left( 1\pm 1\right)
/2|).  \label{B.5}
\end{eqnarray}

In order to estimate the maximal deviations, $x_{\max }$, $y_{\max }$, we
suppose that $u\tau _{1,2}^{\max }/t_{1,2}\gg |x_{\max }|$, $|y_{\max }|$
for typical parameters used in calculating ${\rm G}(\Delta )$ such that $%
x_{\max }=x_{\tau _{1}^{\max }}$ and $y_{\max }=y_{\tau _{2}^{\max }}$.
Thus, the right-hand sides of Eqs. (\ref{B.4}), (\ref{B.5}) do not depend on 
${\bf x}_{\tau }$ and ${\bf y}_{\tau }$ and they can be easily integrated
with the boundary conditions ${\bf x}_{\tau =0,t_{1}}=0$ and ${\bf y}_{\tau
=0,t_{2}}=0$. For the upper-sign contributions, estimating ${\bf u}w(|{\bf %
u|)|\approx }$ $\ell _{c}\overline{\delta \varepsilon }^{2}$ we transform
the right-hand sides of Eqs. (\ref{B.4}), (\ref{B.5}) into $i\overline{%
\delta \varepsilon }^{2}(|t_{1}|+|t_{2}|)/(2m\hbar \ell _{c})$. Thus, the
result for the maximal deviations is 
\begin{equation}
\left| 
\begin{array}{c}
x_{\max } \\ 
y_{\max }
\end{array}
\right| \simeq \frac{\overline{\delta \varepsilon }^{2}(|t_{1}|+|t_{2}|)}{%
16m\hbar \ell _{c}}\left| 
\begin{array}{c}
t_{1}^{2} \\ 
t_{2}^{2}
\end{array}
\right| .  \label{B.6}
\end{equation}
This satisfies the condition $|x_{\max }|$, $|y_{\max }|\ll \ell _{c}$ for
the upper-sign contributions because $t_{1,2}$ are estimated as $\hbar
/\varepsilon _{\scriptscriptstyle F}$. Indeed, then we have $|x_{\max
}|/\ell _{c}$ and $|y_{\max }|/\ell _{c}\approx (\overline{\delta
\varepsilon }/\varepsilon _{\scriptscriptstyle F})^{2}\times (\varepsilon
_{c}/4\varepsilon _{\scriptscriptstyle F})\ll 1$, where the characteristic
energy $\varepsilon _{c}=(\hbar /\ell _{c})^{2}/2m\ll \varepsilon _{%
\scriptscriptstyle F}$.

A more careful consideration is necessary for the lower-sign contributions
when $|t_{1}-t_{2}|\alt\hbar /\varepsilon _{\scriptscriptstyle F}$, such
that $t_{1,2}\approx t$. Thus, Eq. (\ref{B.4}) can be rewritten as 
\begin{equation}
\ddot{{\bf x}}_{\tau }\approx \frac{i{\bf u}\overline{\delta \varepsilon }%
^{2}t}{m\hbar \ell _{c}^{2}}\int_{0}^{1}dz^{\prime }(z-z^{\prime })\exp %
\left[ -u^{2}(z^{\prime }-z)^{2}/\ell _{c}^{2}\right] ,  \label{B.7}
\end{equation}
where $z=\tau /t$. Here, from Eq. (\ref{B.5}), it follows that $\ddot{{\bf y}%
}_{\tau }\approx 0$, and taking into account the boundary conditions, ${\bf y%
}_{\tau =0,t_{2}}=0$, we are led to ${\bf y}_{\tau }\approx 0$. It is easy
to solve Eq. (\ref{B.7}) taking into account in its right hand side that for
the lower-sign contributions $t\gg \tau $, as it was shown in Sec. IV, due
to the fact that $\varepsilon _{\scriptscriptstyle F}\gg \gamma _{%
\scriptscriptstyle eff}$. Then solving Eq. (\ref{B.7}), with boundary
conditions ${\bf x}_{\tau =0,t_{1}}=0$, we obtain the maximal deviation $%
|x_{\max }|$ as 
\begin{equation}
\left| x_{\max }\right| \approx \frac{\overline{\delta \varepsilon }^{2}t^{3}%
}{16m\hbar \ell _{c}}F_{x}(u/\ell _{c}),  \label{B.8}
\end{equation}
where $F_{x}(z)=[1-\exp (-z^{2})]/z$ and $F_{x}(z)\approx z$, for $z^{2}\ll
1 $, while $F_{x}(z)\approx 1/z$, for $z^{2}\gg 1$ and the maximum of $%
F_{x}(z)<0.65$ corresponds to $z$ close to the unity. This estimation
practically coincides with Eq. (\ref{B.6}) if $u/\ell _{c}\sim 1$, while the
maximal deviations are smaller then the results in Eq. (\ref{B.6}) both for $%
u/\ell _{c}\ll 1$ and $u/\ell _{c}\gg 1$.

\bigskip $^{a}$On leave from: Institute of Semiconductor Physics, Kiev,
National Academy of Sciences of Ukraine, 252650, Ukraine

\bigskip \newpage 

\begin{center}
FIGURE CAPTIONS
\end{center}

Fig. 1. a) Spatial variations of the energy levels in $l$- and $r$-QWs along
the $x$ direction without screening (dotted curves) and with screening
(solid curves); b) Band diagram of DQWs, along the $z$ direction, with
nonideal heterointerfaces shown by dashed lines.

\medskip

Fig. 2. Modified lineshapes {\rm G}$(\Delta ),$ taken from Eq. (\ref{28}),
normalized by ${\rm G}_{L}(0)=2(eT)^{2}\rho _{2D}/\hbar \gamma $, when
nonlocal effects are negligible, for different contributions of short-range
scattering, characterized by the phenomenological broadening parameter $%
\gamma $, and non-screened large-scale disorder, characterized by $\overline{%
\delta \varepsilon }$ and $\ell _{c}$. The solid, dashed, dotted,
dot-dashed, and dot-dot-dashed curves correspond to $\overline{\delta
\varepsilon }/\gamma =0.3,$ $0.6,$ $1,$ $3$, and $6$, respectively. The
solid curve is almost a Lorentzian lineshape, while the dot-dot-dashed curve
is very close a Gaussian curve.

\medskip

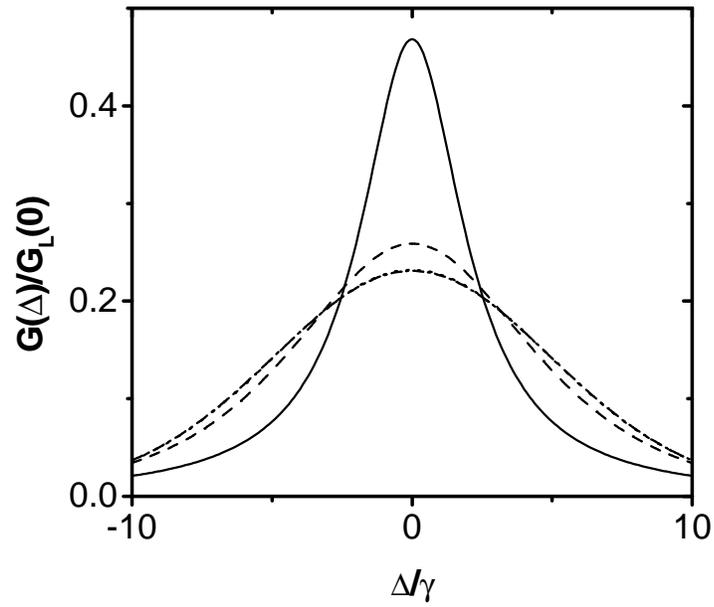
Fig. 3. Modified lineshape ${\rm G}(\Delta )$, calculated from Eq. (\ref{26}%
), when nonlocal effects are taken into account, for $\overline{\delta
\varepsilon }/\gamma =4.6$ and different values of $\hbar v_{F}/\ell
_{c}\gamma $. The solid, dashed, dotted, and dot-dashed curves correspond to 
$\hbar v_{F}/\ell _{c}\gamma =15,3.5,0.7$, and $0.2$, respectively and
represents, for instance, the increase of $\ell _{c}$. Note that the
lineshape evolves practically from a Lorentzian (solid curve), given by Eq. (%
\ref{29}), to a Gaussian one (dot-dashed curve).

\medskip

Fig. 4. Dimensionless tunnel resonance width $\gamma _{eff}/\gamma $ as
function of $\hbar v_{F}/\ell _{c}\gamma $, calculated from Eq. (\ref{26}).
The solid curves from top to bottom correspond to $\overline{\delta
\varepsilon }/\gamma =4.6,$ $2.3,$ $1.5,$ $1.1,$ $0.8$, and $0.3$. For $%
\hbar v_{F}/\ell _{c}\gamma =0$, nonlocal effects are negligible.

\medskip

Fig. 5. Modified lineshape ${\rm G}(\Delta ),$ for data of sample B of Ref. 
\cite{5}, calculated from Eq. (\ref{26}), in which temperature effects were
included. The solid, dashed, dotted and dot-dashed curves correspond to
temperatures $0.7$, $5$, $7$, and $10$ K, respectively. In ${\rm G}_{L}(0)$
it was used $\gamma =0.2$ meV corresponding to the solid curve.


\begin{references}
\bibitem{14}  Notice that the general form of Eq. (\ref{28}) may be obtained
by averaging the Lorentz peak with a random level splitting $\Delta _{{\bf x}%
}$. The essential point here is the condition of the local regime, given by
Eq. (\ref{27}).
\bibitem{16}  For two-side variations of heterointerfaces, as in Eq. (\ref
{(A.6)}), we obtain $\ell_{c}=350$ \AA , $700$ \AA\ and $500$ \AA\ for
samples A, B and C respectively. Because in Ref. \cite{5} it is reported
that $200$ \AA\ QWs are separated by the tunnel barriers with $175$ \AA\ to $%
350$ \AA\ (even though the real values are not given samples for A, B and C)
and considering the results of Ref. \cite{15}, we have assumed here only
one-side fluctuations of the QW for samples in Ref. \cite{5}.
\end{references}
\end{document}